\documentclass[twocolumn]{aastex62}
\usepackage{amsmath}
\renewcommand{\vec}[1]{\boldsymbol{#1}}
\usepackage{float}
\graphicspath{{./}{figures/}}

\shorttitle{An Early Solar Binary Companion}
\shortauthors{Siraj \& Loeb}

\begin{document}
\title{The Case for an Early Solar Binary Companion}

\email{amir.siraj@cfa.harvard.edu, aloeb@cfa.harvard.edu}

\author{Amir Siraj}
\affil{Department of Astronomy, Harvard University, 60 Garden Street, Cambridge, MA 02138, USA}

\author{Abraham Loeb}
\affiliation{Department of Astronomy, Harvard University, 60 Garden Street, Cambridge, MA 02138, USA}




\begin{abstract}

We show that an equal-mass, temporary binary companion to the Sun in the solar birth cluster at a separation of $\sim 10^3 \; \mathrm{\; AU}$ would have increased the likelihood of forming the observed population of outer Oort cloud objects and of capturing Planet Nine. In particular, the discovery of a captured origin for Planet Nine would favor our binary model by an order of magnitude relative to a lone stellar history. Our model predicts an overabundance of dwarf planets, discoverable by LSST, with similar orbits to Planet Nine, which would result from capture by the stellar binary.

\end{abstract}



\section{Introduction}

Simulations of outer Oort cloud (OOC)\footnote{Defined here as the collection of solar system bodies with semimajor axes of $\sim 10^4 - 10^5 \mathrm{\; AU}$, and with orbits decoupled from Neptune.} formation in the Solar system  \citep{2004come.book..153D, 2008Icar..197..221K, 2010A&A...516A..72B} have difficulties reproducing the observed ratio between scattered disk) (SD)\footnote{Defined here as the collection of solar system bodies with semimajor axes of $\lesssim 10^3 \mathrm{\; AU}$, and with orbits controlled by Neptune.} and OOC objects \citep{1997Sci...276.1670D, 2008AJ....136.1079L} As a result, the origin of the OOC is an unsolved puzzle. Scenarios positing that the formation of the OOC occurred in the stellar birth cluster of the Sun tend to rely on drag from dense cluster gas \citep{2000Icar..145..580F, 2006Icar..184...59B,  2008Icar..197..221K, 2010Sci...329..187L, 2012Icar..217....1B}, a factor that hinders the scattering of comets to large distances, reducing the plausibility of the explanations \citep{2007Icar..191..413B, 2013Icar..225...40B}. There are $N_{OC} \sim 7.6 \pm 3.3 \times 10^{10}$ OOC bodies and $N_{SD} \sim 1.7^{+ 3.0}_{- 0.9} \times 10^9$ SD bodies with diameters of $D > 2.3 \mathrm{\; km}$ \citep{2013Icar..225...40B, 2017A&A...598A.110R}. Simulations of OOC formation due to a dynamical instability in the solar system result in an OOC/SD ratio of $N_{OC}/N_{SD} \sim 12 \pm 1$, which is in tension, but not incompatible with, the observed ratio \citep{2013Icar..225...40B}.

Separately, clustering of extreme trans-Neptunian objects (ETNOs) in the outer solar system suggest the possible existence of a planet, labeled Planet Nine, at a distance of $\sim 500 \mathrm{\; AU}$ from the Sun \citep{2016ApJ...824L..23B, 2019PhR...805....1B}. \cite{2020arXiv200400037Z} argued that Planet Nine may not exist, and its observed gravitational effects could potentially be caused by an unobserved ring of small bodies in the outer solar system. There also exists the possibility that the clustering is a statistical fluke \citep{2020AJ....159..285C}. The origin of Planet Nine, if it exists, is a second unsolved puzzle in the outskirts of the Solar system. Possible solutions include \citep{2019PhR...805....1B} formation amongst the giant planets followed by scattering and orbital circularization \citep{2006Icar..184...59B, 2012Icar..217....1B, 2016ApJ...823L...3L}, and capture in the solar birth cluster \citep{2016ApJ...823L...3L, 2016MNRAS.460L.109M, 2017MNRAS.472L..75P}.

Interestingly, stellar binary systems are capable of capturing background objects via three-body processes \citep{1975MNRAS.173..729H, 1983Obs...103....1V}, leading to capture rates that are enhanced relative to lone stars \citep{2018ApJ...868L..12G, 2020arXiv200102235S}. Current binary companions to the Sun were previously considered \citep{2005EM&P...97..459M, 2010MNRAS.407L..99M}, and subsequently ruled out \citep{2014ApJ...781....4L}. Here, we consider a temporary binary companion to the Sun that could have existed only in the solar birth cluster, and explore the plausibility and implications of such a possibility for both the formation of the OOC and the capture of Planet Nine.

Our discussion is structured as follows. In Section \ref{sec:p}, we explore the plausibility of a binary companion to the Sun in the solar birth cluster. In Section \ref{sec:oc}, we investigate the effects of an early binary on the formation of the OOC. In Section \ref{sec:cr}, we consider the implications of the binary model for the Planet Nine capture cross-section and use the likelihood of the binary configuration considered to estimate the overall merits of binary model if a captured origin for Planet Nine is verified. Finally, Section \ref{sec:d} summarizes the key implications of our model. 

\section{Plausibility}
\label{sec:p}

\begin{figure}[H]
\smallskip
  \centering
  \includegraphics[width=0.9\linewidth]{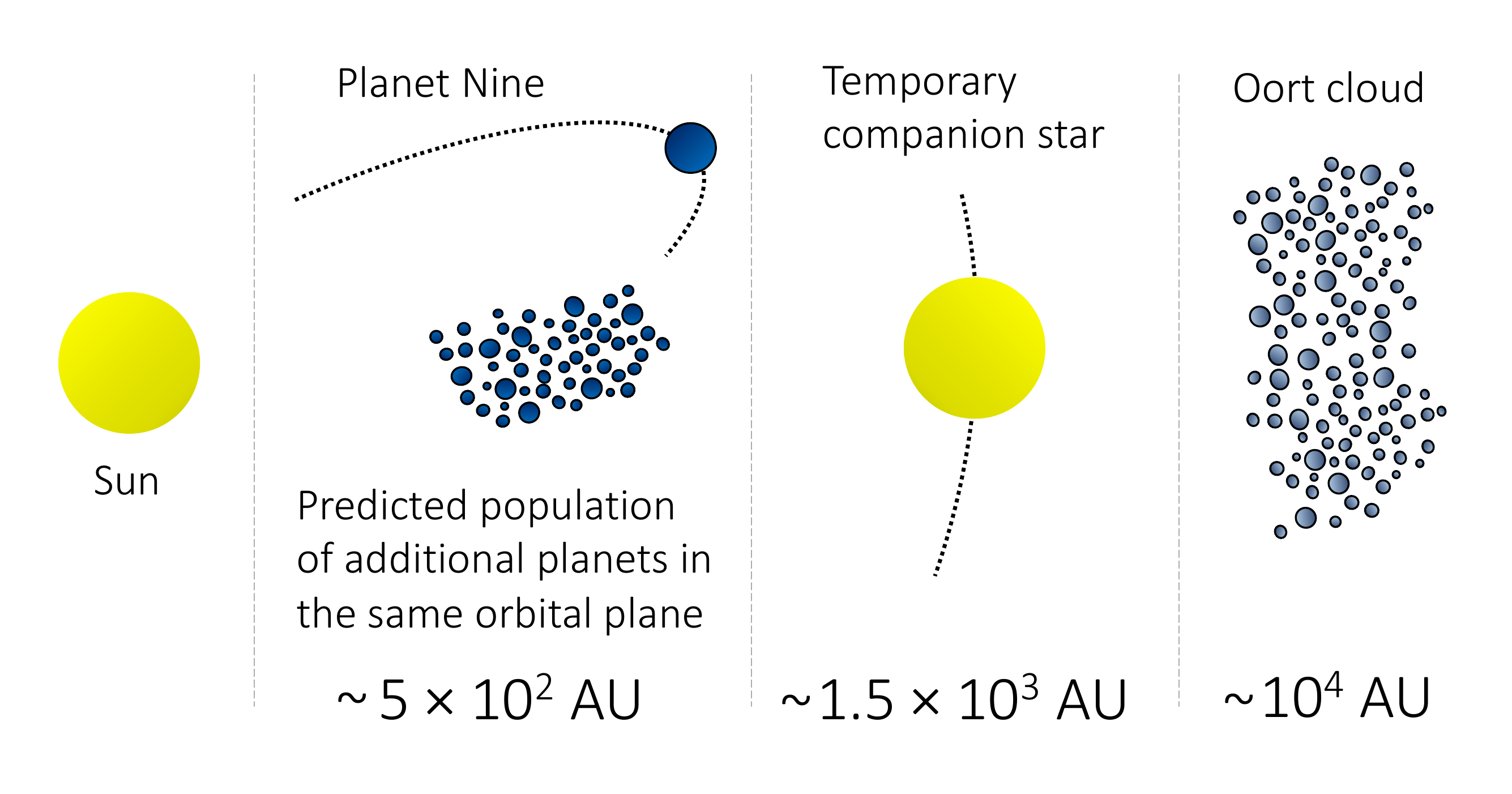}
    \caption{Sketch of scenario considered here (not to scale).
}
    \label{fig:sketch}
\end{figure}

The orbit of Planet Nine would be stable in an equal-mass binary if the binary separation were a factor of $\sim 3$ larger than that of Planet Nine (Fig. 4, \citealt{2020AJ....159...80Q}). Since the semi-major axis of Planet Nine is likely $\sim 500 \; \mathrm{AU}$, we consider an equal-mass binary with a separation of $a \sim 1500 \mathrm{\; AU}$, at which a solar binary could have plausibly been born and survived the protostellar phase \citep{2008AJ....135.2526C}, although more research may be necessary \citep{2017MNRAS.469.3881S}. Figure \ref{fig:sketch} shows a sketch of the scenario considered here. Separations of $\gtrsim 1500 \mathrm{\; AU}$ are possible, but they would reduce both the capture cross-section, which scales as $a^{-1}$, and the lifetime in the birth cluster, which scales as $a^{-1/2}$.

The orbits of the planets in the Solar system would be unaffected by Kozai-Lidov oscillations from such a binary partner (see Table 1 in \citealt{1997AJ....113.1915I}; extrapolated using the $b_{\star}^3 / m_{\star}^3$ relation).

Since the ejection probability for a body at a separation of $\sim 500 \mathrm{\; AU}$ over the lifetime of the solar birth cluster is $\sim 0.3$ \citep{2019PhR...805....1B}, and orbital speed scales as $a^{-1/2}$ while the distribution of $\Delta v$ impulses is the same at any point in space where gravitational focusing is not significant, the probability of ejection for an object with a separation of $a \sim 1500 \mathrm{\; AU}$ is $f_e \sim 0.5$ ($f_e \sim 1$ for $a \sim 6000 \mathrm{\; AU}$), which is consistent with the fact that no solar-mass binary companion is presently observed.

Additionally, since tidal force scales as the cube of distance, Planet Nine with a perihelion of $\sim 250 \; \mathrm{AU}$ and mass $\sim 5 - 10 \; M_{\oplus}$ acting over a timescale comparable to the age of the solar system $\sim 4.5 \mathrm{\; Gyr}$, would have a comparable effect on solar obliquity as a binary stellar companion \citep{2012Natur.491..418B} with a perihelion of $\gtrsim 1500 \; \mathrm{AU}$, and a mass of $\sim M_{\odot}$, acting over the lifetime of the solar birth cluster $\sim 0.1 \mathrm{\; Gyr}$ \citep{2016AJ....152..126B}. Furthermore, a binary stellar companion could potentially produce the observed \citep{2016ApJ...833L...3B, 2016ApJ...827L..24C} high-inclination Centaurs.

Furthermore, the evidence in the distribution of long-period comets for a Jupiter-mass solar companion at a distance of $\sim 10^4 \mathrm{\; AU}$ \citep{2011Icar..211..926M}, which was ruled out \citep{2014ApJ...781....4L}, could be consistent with the effects of an equal-mass binary companion at $\sim 1500 \mathrm{\; AU}$ acting over a timescale of $\sim 0.1 \mathrm{\; Gyr}$. The impulse delivered by a binary companion to objects in the OOC scales inversely with the square of the distance between the companion and the OOC objects \citep{2009NewA...14..166B}. While the impulse also scales inversely with the orbital speed of the binary companion, this effect is compensated for by the orbital period scaling with the orbital speed. The total magnitude impulses delivered by the hypothetical Jupiter-mass companion at $\sim 10^4 \mathrm{\; AU}$ to OOC objects at similar separations from the Sun would be comparable to those provided equal-mass binary companion at a separation of $\sim 1500 \mathrm{\; AU}$ in the solar birth cluster. The overall structure of the OOC, however, is in a steady-state with little dependence on initial conditions \citep{2017Icar..292..218F}, which is encouraging for the binary capture model of OOC objects.

Finally, we note that the product of velocity dispersion, stellar density, and lifetime, for the solar birth cluster and field, respectively, are comparable to order unity, implying that the minimum impact parameter of a stellar encounter relative to the Sun is similar between the birth cluster and the field. Since the impulse approximation dictates that the velocity shift, $\Delta v$, imparted to an orbiting body by a perturber is inversely proportional to the relative encounter speed, the dissociation of the stellar companion from the Sun is more likely to have taken place in the birth cluster than the field by more than an order of magnitude, due to the difference in velocity dispersion between the birth cluster and the field.


\section{Outer Oort cloud formation}
\label{sec:oc}

Next, we quantify the expected abundance of OOC objects for the model considered here, in order to compare to both observations and other models. 2I/Borisov is the only confirmed interstellar comet \citep{2020NatAs...4...53G}.\footnote{We do not consider 1I/`Oumuamua to be a traditional comet given the lack of observed outgassing \citep{2018Natur.559..223M, 2018AJ....156..261T}.} The number density of Borisov-like objects is $n_B \sim 8.8 \times 10^{-3} \; \mathrm{AU^{-3}}$ \citep{2019ApJ...886L..29J}. Since the local number density of stars is $n_f \sim 0.14 \; \mathrm{pc^{-3}}$, we estimate that each star produces $\sim 5.5 \times 10^{14}$ Borisov-like objects. The nucleus of Borisov likely had a radius of $0.4 - 1 \mathrm{\; km}$, with an absolute lower bound on radius of $\sim 0.35 \mathrm{\; km}$ \citep{2020ApJ...888L..23J} which we adopt here as a conservative measure, along with a cumulative size distribution with a power-law index $-3$, corresponding to equal mass per logarithmic bin, as justified by the size distribution of interstellar objects \citep{2019arXiv190603270S}. The number of $D \gtrsim 2.3 \mathrm{\; km}$ interstellar comets produced per star is thereby estimated to be, $\sim 1.6 \times 10^{13}$. The total capture cross-section\footnote{For marginally bound objects with $E \sim 0$, where E is defined after the companion and most objects leave the system.} for a solar-mass binary with separation $\sim 1500 \mathrm{\; AU}$ for objects with a velocity dispersion of $v \sim 1 \mathrm{\; \mathrm{km \; s^{-1}}}$ is $\sigma \sim 1.6 \times 10^{6} \mathrm{\; AU}$ \citep{1975MNRAS.173..729H, 1983Obs...103....1V}, and we adopt a cluster stellar density of $n_c \sim 100 \; \mathrm{pc^{-3}}$ and lifetime of $\tau \sim 10^8 \mathrm{\; yr}$ \citep{2010ARA&A..48...47A}, which is consistent with the limit set by the observed inclination of the cold classical Kuiper belt \citep{2020AJ....159..101B}. The fraction of the interstellar comets produced per star captured by such a binary over the lifetime of the solar birth cluster is $(\sigma v \tau n_c) \sim 40 \%$. As a result, the number of captured objects over the lifetime of the birth cluster is expected to be $\sim 6.4 \times 10^{12}$.

The closest stellar encounters have the greatest effects on erosion of OOCs \citep{2018MNRAS.473.5432H}, so here we focus on the closest stellar encounter to the solar system over the cluster lifetime, and assume that this encounter unbound the stellar binary. Ignoring the gas-rich initial period lasting $\sim 1 \mathrm{\; Myr}$, the impact parameter of the closest stellar encounter over the cluster lifetime is estimated to be $b \sim (n_c \tau v)^{-1/2} \sim 2 \times 10^3 \mathrm{\; AU}$. We adopt a distance an order of magnitude larger than this impact parameter as the fiducial separation between the Sun and the outer OOC, $r \sim 2 \times 10^4 \mathrm{\; AU}$. The mass of the perturbing star is assumed to be, $M_p \sim 0.1 \; M_{\odot}$. 

The impulse approximation, which holds since $v_p \gg \sqrt{2GM_{\odot}/r}$, for the velocity kick of an OOC object relative to the Sun as a result of a stellar perturbation \citep{2009NewA...14..166B} gives,

\begin{equation}
    \Delta \vec{v} = \frac{2 G M_p r}{b^2 v} \; [\hat{\vec{r}} - 3 \hat{\vec{b}} (\hat{\vec{r}} \cdot \hat{\vec{b}}) - \hat{\vec{v}}^p (\hat{\vec{r}} \cdot \hat{\vec{v}}^p)] \; \; ,
\end{equation}
where $\hat{\vec{r}}$ is the vector from the Sun to the OOC object, $\hat{\vec{b}}$ is the impact parameter vector from the Sun to the closest approach of the perturber, and $\hat{\vec{v}}^p$ is the velocity vector of the perturber.

For simplicity, we consider a model in which the trajectory of the perturber is normal to the orbital plane of the binary, in which case OOC objects with position vectors aligned or anti-aligned with the perturber trajectory would receive no velocity kick relative to the Sun, meaning that they remain bound. In particular, the condition for remaining bound post-perturbation is $\Delta v \lesssim \sqrt{G M_{\odot} / r}$. For the conventions described above, any OOC object within $\sim 14^{\circ}$ of the perturber's trajectory should remain bound to the Sun. The infinitesimal element of solid angle is $d[\sin \theta]$, and since the average value of $\sin \theta$ over the range of possible perturber trajectory angles is $(2/\pi)$, we use the small angle approximation to apply a correction factor of $(2/\pi)$ to the range of angular separations for which OOC objects survive, resulting in a value of $\sim 9^{\circ}$ for a typical perturber trajectory. The area covered by points within $\lesssim 9^{\circ}$ of a diameter vector of a sphere is $\sim 1 \%$ of the surface area. We thereby estimate that $\sim 99 \%$ of OOC objects are lost due to the stellar encounter that unbinds the binary, resulting in $\sim 8 \times 10^{10}$ surviving objects at the end of the birth cluster lifetime.

The disruptions of OOC orbits by additional passing stars from the birth cluster are relatively insignificant since, as a result of the only the closest stellar encounter, the ejection fraction for OOC objects is $f_e \sim 1$. If we considered the next logarithmic bin of impact parameters, reasoning that the combination of the impulse approximation giving $\Delta \vec{v} \propto b^2$ and the fact that $P(b) \propto b^2$ would yield a comparable cumulative $\Delta v$, the impulse approximation would break down since $v_p \sim \sqrt{2 G M_{\odot} / r}$, necessitating that we rely on the results of direct simulations like those of \cite{2018MNRAS.473.5432H}, which show that the closest encounters dominate the loss of comets. We note that the survival of $\sim 1 \%$ of the objects captured during the lifetime of the birth cluster the outer OOC is consistent with the finding that $35\% - 75\%$ of objects survive over the lifetime of the solar system excluding the birth cluster \citep{2018MNRAS.473.5432H}, since the total numbers of stellar encounters inside and outside of the birth cluster are comparable, and the velocity kick per encounter in the cluster is $\sim 20$ times larger than in the field.

\begin{figure}
  \centering
  \includegraphics[width=0.9\linewidth]{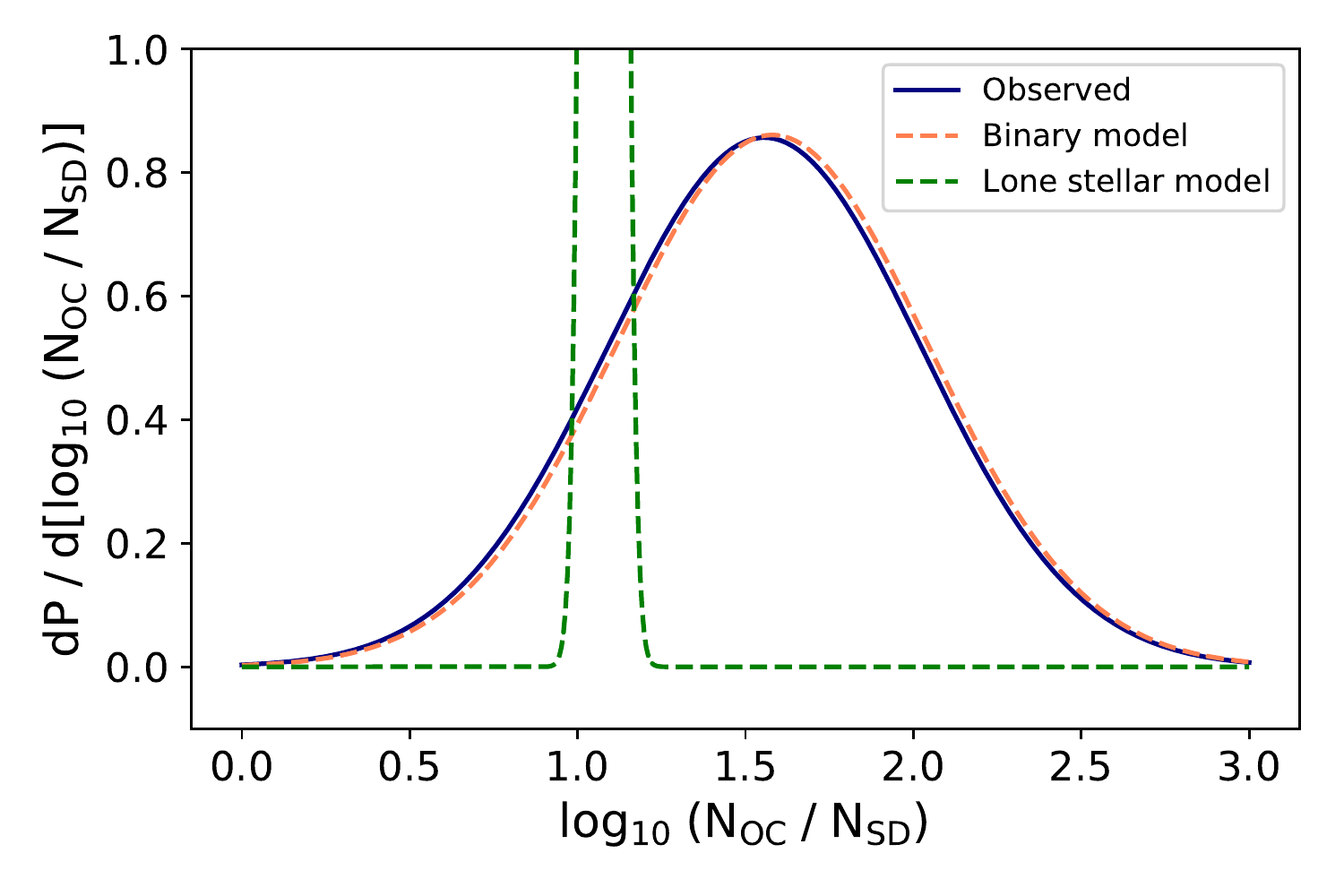}
    \caption{The normalized probability distributions of the ratio between OOC and SD objects for the binary model described here and for the lone stellar model \citep{2013Icar..225...40B}, with the observed ratio \citep{2013Icar..225...40B} shown for reference.
}
    \label{fig:mc}
\end{figure}

Propagating the aforementioned uncertainty on the size of Borisov, we estimate that a binary would result in an OOC with $N_{OC} \sim 8 \pm 3.4 \times 10^{10}$ comets with $D > 2.3 \mathrm{\; km}$, which is in excellent agreement with the observed value of $N_{OC} \sim 7.6 \pm 3.3 \times 10^{10}$. Based on these calculations, we used a Monte Carlo simulation, the results of which are shown in Figure \ref{fig:mc}, to quantify the goodness-of-fit of our model versus that of \cite{2013Icar..225...40B} relative to the observations, and found that the overlapping coefficient for the former is $\sim 5$ times greater than the latter, implying that based upon the current understanding of the OOC, our binary model increases the chances of forming the observed number of OOC objects by a factor of $\sim 5$ relative to the lone stellar model.

\section{Planet Nine \& Overall Likelihood}
\label{sec:cr}

We now consider the implications of the binary model for the Planet Nine capture cross-section and evaluate the binary model conditional upon a captured origin for Planet Nine being verified. The Planet Nine capture-cross-section for a binary stellar system is \citep{1975MNRAS.173..729H, 1983Obs...103....1V},

\begin{equation}
\begin{aligned}
    \sigma \sim 2 \times 10^{5} \mathrm{\; AU^2} & \left( \frac{a}{1500 \; \mathrm{AU}} \right)^{-1} \left( \frac{m}{M_{\odot}} \right)^2 \left( \frac{m + M_{\odot}}{2 M_{\odot}} \right)^{-1}  \\ & \left( \frac{v}{\mathrm{1 \; km \; s^{-1}}} \right)^{-1} \left( \frac{\sqrt{v^2 + v_c^2}}{\mathrm{\sqrt{2} \; km \; s^{-1}}} \right)^{-6} \; \; ,
\end{aligned}
\end{equation}
where $a$ is the semi-major axis of the binary, $m$ is the mass of the Sun's binary companion, $v$ is the typical encounter velocity in the solar birth cluster, and $v_c$ is the orbital speed of the captured orbit of Planet Nine. This cross-section is a factor of $\sim 20$ times greater than the Planet Nine capture-cross section for a lone solar-type star (Table 1, \citealt{2016ApJ...823L...3L}). The binary and lone capture cross-sections would undergo the same enhancements when considering the capture of a planet bound to another star.

We now consider the likelihood of the binary configuration considered here and how this forms the overall statistical argument. The fraction of solar-type stars with roughly equal-mass binary companions is $f_m \sim 0.25$, since there is an overabundance of observed equal-mass binary companions relative to lower masses (Fig. 16, \citealt{2010ApJS..190....1R, 2019MNRAS.489.5822E}). As explained in Section \ref{sec:p}, we only consider binary companions at separations $a \gtrsim 1500 \mathrm{\; AU}$. The probability for a binary partner with a separation of $\gtrsim 1500 \; \mathrm{AU}$ relative to one with a separation of $\gtrsim 500 \; \mathrm{AU}$ is
$f_a \sim 0.6$ (Fig. 16, \citealt{2010ApJS..190....1R}). Since $\gtrsim 50 \%$ of solar-type stars are members of binaries \citep{2010ApJS..190....1R, 2013ARA&A..51..269D}, the likelihood of the binary configuration described here is $f_{m} f_{a} f_{e} \sim 10 \%$. We note that these values are primarily based upon observations of binaries in the field and therefore may be conservative for binaries in clusters.

Since the binary model improves the likelihood of the observed OOC population by a factor of $\sim 5$ and the capture of a putative Planet Nine by a factor of $\sim 20$, whereas the binary configuration considered here applies to $\sim 10\%$ of solar-type stars, we find that the discovery of a captured Planet Nine would result in the binary model being favored by an order of magnitude relative to the conventional lone stellar model.

\section{Discussion}
\label{sec:d}

We propose that an equal-mass binary companion to the Sun in the solar birth cluster at a separation of $\sim 10^3 \mathrm{\; AU}$ would explain the formation of the observed population of OOC objects and the putative existence of Planet Nine. Separations greater than the fiducial example given here, $a \sim 1500 \; \mathrm{AU}$, are entirely plausible; the capture cross-section would simply scale as $a^{-1}$ and the likelihood of ejection in the birth cluster as $a^{1/2}$, up to a maximum of $a \sim 6000 \; \mathrm{AU}$ since the chance of ejection in the birth cluster would then be of order unity. If Planet Nine is discovered, evidence of a captured origin, as opposed to formation within the Solar system, could potentially come from a cloud of objects with associated orbits \citep{2016MNRAS.460L.109M}. Accounting for the likelihood of the binary configuration considered here, the discovery of a captured Planet Nine would favor our binary model by a factor of $\sim 10$, when the increased likelihoods of both forming the OOC and capturing Planet Nine are considered.

The specific smoking gun for our binary model will be a significant overabundance of dwarf planets with similar orbits to Planet Nine, since the the capture cross-section for such objects would have been a factor of $\sim 20$ larger than implied by the conventional lone stellar model, and given that orbits situated closer to the proposed binary than Planet Nine would be unstable \citep{2020AJ....159...80Q}. These objects could potentially be detected by the Legacy Survey of Space and Time (LSST)\footnote{https://www.lsst.org/} on the Vera C. Rubin Observatory. In addition, since the binary model would bring the likelihood of Planet Nine capture in the solar birth cluster near unity, the existence of a captured planet in addition to Planet Nine would be probable. Detailed modeling of the effects of a binary on long-period comets, the solar obliquity, and ETNOs will allow for the development of additional tests.

\section*{Acknowledgements}
\vspace{0.1in} 
We thank Fred Adams, Gongjie Li, and Konstantin Batygin for helpful comments on the manuscript. This work was supported in part by the Origins of Life Summer Undergraduate Research Prize Award and a grant from the Breakthrough Prize Foundation.





\newpage 

\bibliography{bib}{}
\bibliographystyle{aasjournal}



\end{document}